\documentclass{article}
\usepackage{spconf,amsmath,graphicx}
\usepackage{color}
\usepackage{multirow}
\usepackage{lipsum}  
\usepackage{booktabs,siunitx}
\usepackage{colortbl}
\usepackage{amssymb}
\usepackage{pifont}
\usepackage{amsfonts}
\usepackage{tabularx}
\usepackage{array, makecell}
\usepackage{xcolor}
\usepackage{graphicx}
\usepackage{kotex}
\usepackage{comment}
\usepackage{soul}
\usepackage{url}
\usepackage{amsmath}
\usepackage{kotex}

\title{Textless Speech-to-Music Retrieval Using Emotion Similarity}

\name{$^\flat$SeungHeon Doh, $^\natural$Minz Won, $^\sharp$Keunwoo Choi, $^\flat$Juhan Nam} 
\address{$^\flat$Graduate School of Culture Technology, KAIST, South Korea \\ $^\natural$ByteDance, USA \\ $^\sharp$Gaudio Lab, South Korea
}
\begin{document}
\ninept
\maketitle
\begin{abstract}
We introduce a framework that recommends music based on the emotions of speech. In content creation and daily life, speech contains information about human emotions, which can be enhanced by music. Our framework focuses on a cross-domain retrieval system to bridge the gap between speech and music via emotion labels. We explore different speech representations and report their impact on different speech types, including acting voice and wake-up words. We also propose an emotion similarity regularization term in cross-domain retrieval tasks. By incorporating the regularization term into training, similar speech-and-music pairs in the emotion space are closer in the joint embedding space. Our comprehensive experimental results show that the proposed model is effective in textless speech-to-music retrieval. 

\end{abstract}

\section{Introduction}

\begin{figure}[!t]
\centering
\includegraphics[width= 0.9\columnwidth]{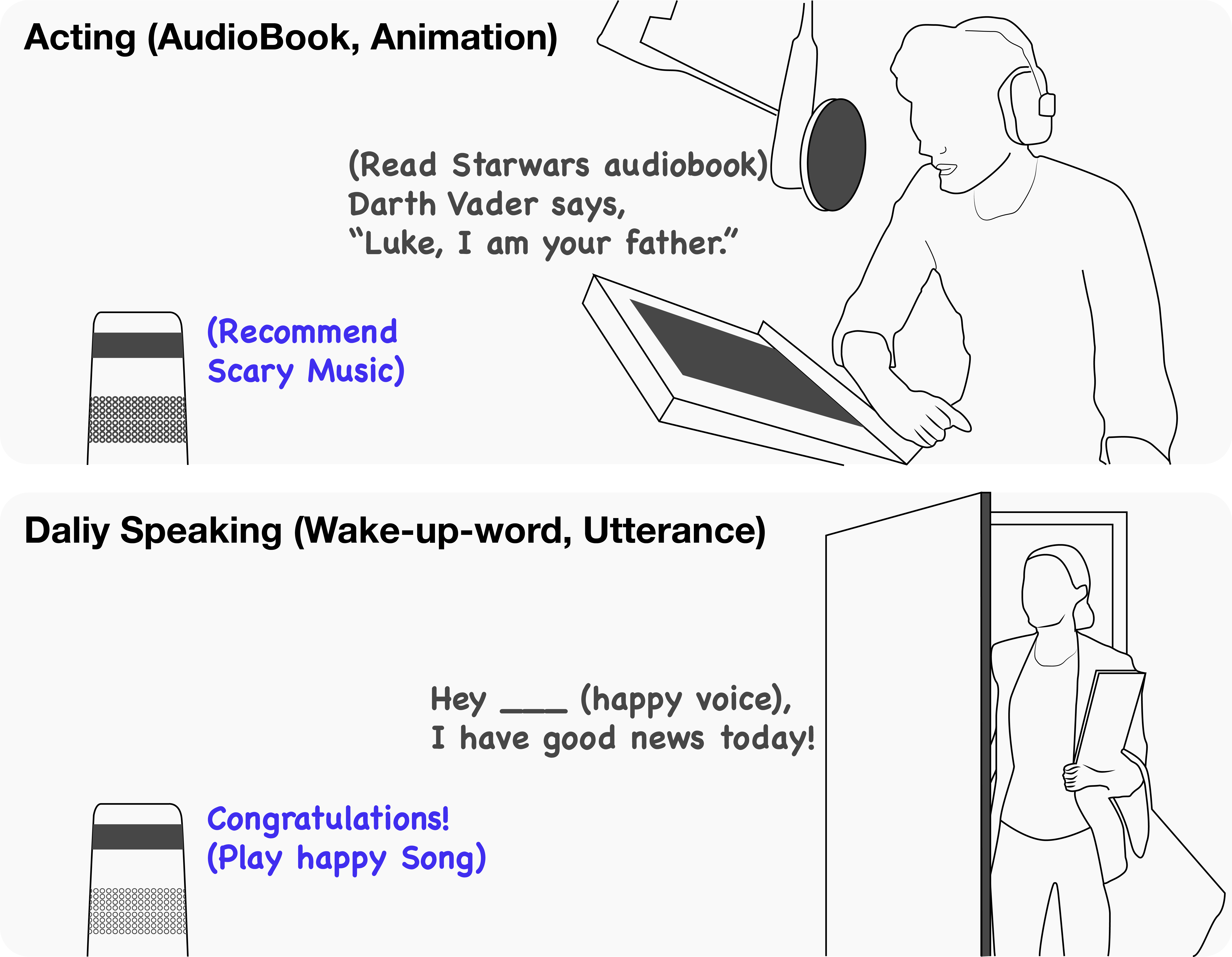}
\caption{User scenario of speech-to-music retrieval with a smart speaker. 
}
\label{fig:user}
\vspace{-4mm}
\end{figure}

In multimedia content, emotions are delivered firstly through speech, facial expressions, and gestures. The presented emotions are often reinforced when they are matched with suitable music. For example, if a movie scene is about a romantic conversation, the mood will be enhanced with a romantic soundtrack (illustrated in Figure~\ref{fig:user}). However, this cross-domain matching process, i.e., speech emotion recognition and music search, is time-consuming and requires domain expertise.

In order to tackle such retrieval problems, previous studies formalize the task as a cross-modal retrieval~\cite{wang2017adversarial} -- e.g., music video-to-music~\cite{Hong2018CBVMRCV}, user video-to-music~\cite{li2019query, suris2022s}, and storytelling text-to-music~\cite{won2021emotion}. A common approach for cross-domain retrieval is multimodal representation learning which transforms the data samples from different modalities into a joint embedding space. The joint embedding space is optimized to align data distributions of heterogeneous modalities. The alignment problem was solved by unsupervised, pairwise-based, rank-based, and supervised methods in \cite{wang2016comprehensive}, where the methods leverage co-occurrence data, pairwise data, rank lists, and annotated labels, respectively. However, speech and music cannot co-occur organically -- they are only intentionally made to co-occur -- hence we do not have an organically paired dataset of speech and music. A possible solution to rely on unpaired, separately existing speech datasets and music datasets is a supervised method that uses the relationship between the labels of each domain.

However, mismatched dataset taxonomies (emotion vocabularies) between domains cause immediate difficulties. This issue has been addressed by emotion-based mapping where each tag is transformed into the most similar emotion tag, e.g., joy (from speech dataset)$\xrightarrow{}$happy (from music data). Won~\textit{et al.}~\cite{won2021emotion} formalized the task of matching storytelling to music as a supervised cross-modal retrieval. They introduced three possible emotion-based mappings: (1) mapping based on the Euclidean distance between emotion labels in the predefined valence-arousal space, (2) the cosine distance between emotion labels in predefined word embedding space, or (3)~direct manual mapping of emotion labels. This emotion-mapping approach allows the model to bridge the modality gap. While such mapping methods provide a data-driven alignment, this may lack information on continuous emotion distribution due to the discrete nature of the most similar mapping. A direct, manual mapping may address the issue of most similar mapping by one-to-many mapping (e.g., joy$\xrightarrow{}$\{happy, excitement\}). The annotation cost increases proportionally with the number of labels.

\begin{figure*}[!t]
\centering
\includegraphics[width=0.93\textwidth]{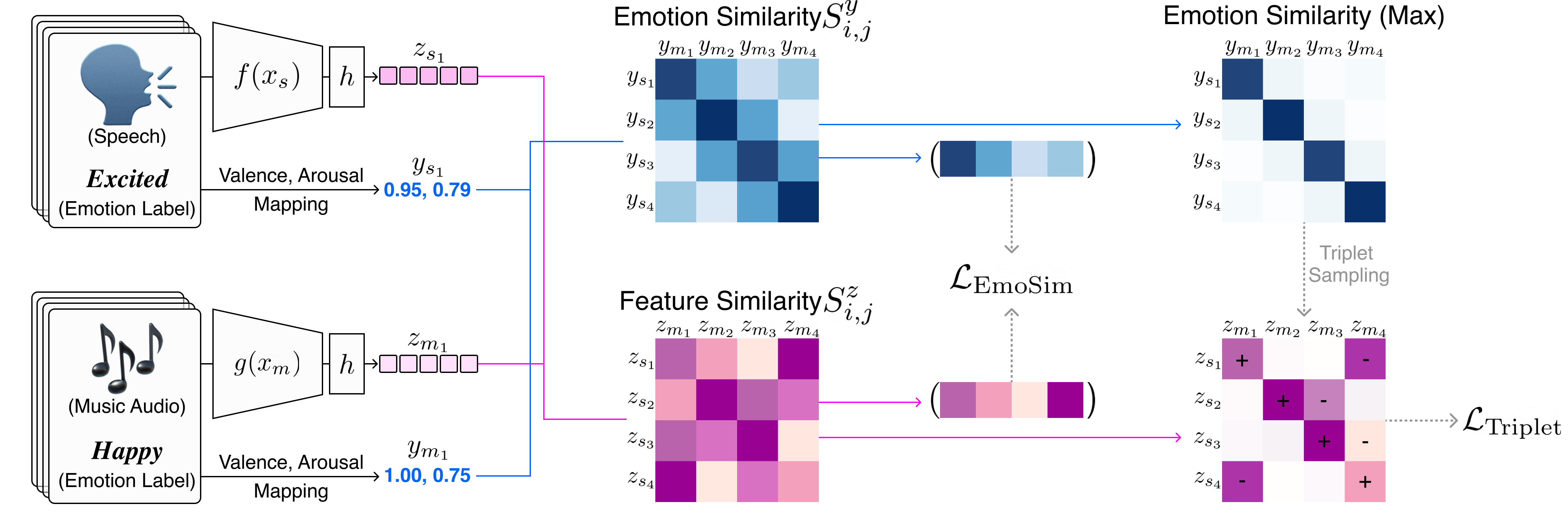}
\caption{The proposed speech-to-music retrieval framework. $S_{y}$ and $S_{z}$ are encoded pairwise similarities of the emotion label and feature, respectively. The goal of the loss function is to encourage items (speech, music) with similar emotion labels to be similar in the feature space.}
\label{fig:2}
\vspace{-3mm}
\end{figure*}

In this work, we formalize matching the speech-to-music task as a supervised cross-domain retrieval problem.\footnote{We use the term \textbf{cross-domain} instead of \textbf{cross-modal} because speech audio and music audio share the same modality.} Inspired by \cite{gong2022ranksim}, we propose an emotion similarity regularization term, which take advantage of continuous emotion similarity information. For example, we guide the network in a way that the feature representation for the happy speech is more similar to the music representation in order: exciting, funny, tensor, sad, angry, and scary music. We explore different speech representations and report their impact on different speech types including acting voice and wake-up words. Experiments show that we can successfully bridge the gap between speech and music to facilitate cross-domain retrieval. We present the code and demo online.\footnote{{https://seungheondoh.github.io/speech-to-music-demo}}



\section{Speech-to-Music Retrieval}
Similar to the previous work~\cite{won2021emotion}, our cross-domain retrieval model comprises two parts: pre-trained encoder networks for feature extraction and projection layers for bridging the domain gap. In the proposed systems, speech is represented by audio, text, and fusion modalities. As shown in Figure~\ref{fig:2}, the architecture takes two different domain data from speech and music. Speech data include audio, transcription text, and emotion labels. Music data include audio and emotion labels. The domain data are processed by pre-trained encoder networks and projection multilayer perceptrons (MLPs). The pre-trained encoder networks extract a domain feature, and the projection MLPs transform the outputs of each domain feature into a joint embedding space.

\subsection{Pre-trained Models for Feature Extraction}
The large-scale pre-trained model revealed promising generalization performance in emotion recognition as a downstream task, for example, music \cite{choi2017transfer, castellon2021codified}, speech \cite{chen2021exploring, cai2021speech, pepino2021emotion}, text-based \cite{adoma2020comparative}.
As shown in Figure \ref{fig:2}, the proposed architecture transforms music, speech into a joint space through pre-trained encoder models. For the music encoder, we use the Music Tagging Transformer~\cite{won2021transformer} which is the current state-of-the-art music tagging model. For the speech audio encoder, we use Wav2vec~2.0~\cite{baevski2020wav2vec} and take the average pooling on its phoneme-level feature to summarize them into utterance-level \cite{cai2021speech}. For the speech text encoder, we use DistilBERT~\cite{sanh2019distilbert}, which is a compact variant of the popular BERT transformer model. For music audio, speech audio, and speech text encoders, we use pre-trained networks fine-tuned through a classification task (See Table \ref{tab:baseline}). For the emotion label, we use valence-arousal representation from VAD Lexicon~\cite{mohammad-2018-obtaining}.

\subsection{Cross-Domain Retrieval}
In this section, we introduce speech-to-music retrieval frameworks (illustrated in Figure~\ref{fig:2}). In the following descriptions, the notations $x_{s}, y_{s}$ and $x_{m}, y_{m}$ represent the speech and music examples, along with their corresponding emotion labels. The notations $f$ and $g$ denote the pre-trained encoders for speech and music, and $h_{s}$ and $h_{m}$ denote the projection MLPs for speech and music, respectively. Each domain input is processed by the corresponding pre-trained encoder $f$ or $g$ and projection MLP $h_{s}$ or $h_{m}$. We denote the two output embeddings as $z_{s} = h_{s}(f(x_{s}))$ and $z_{m} = h_{m}(g(x_{m}))$, respectively. On the cross-domain retrieval model, we freeze the pre-trained encoder $f,g$, and train the projection MLP network $h_{s}, h_{m}$.

We trained our retrieval model using deep metric learning with a triplet loss. The goal of metric learning is to learn an embedding space where inputs from the same semantics are closer to each other than those from different semantics. Given a triplet embedding ($z$, $z^{+}$, $z^{-}$), the metric learning model is optimized to minimize a triplet loss.
\begin{equation}
  \mathcal{L}_{\text{Triplet}}(z, z^{+}, z^{-})=\text{max}\{ 0, D(z,z^{+}) - D(z,z^{-}) + \delta \}
\end{equation}
where $z, z^{+}, z^{-}$ are embedding of anchor, positive and negative sample. $D$ is a cosine distance function and $\delta$ is a predefined margin. In our cross-domain retrieval scenario, the anchor is query speech embedding ($z_{s}$), and positive and negative are music embeddings ($z_{m}^{+},z_{m}^{-}$). Our baseline model is the conventional triplet network that optimizes the loss function $\mathcal{L}$:

\begin{equation}
  \mathcal{L}_{\text{Cross}}= \mathcal{L}_{\text{Triplet}}(z_{s},z_{m}^{+},z_{m}^{-})
\end{equation}

\begin{figure}[!t]
\centering
\includegraphics[width=\columnwidth]{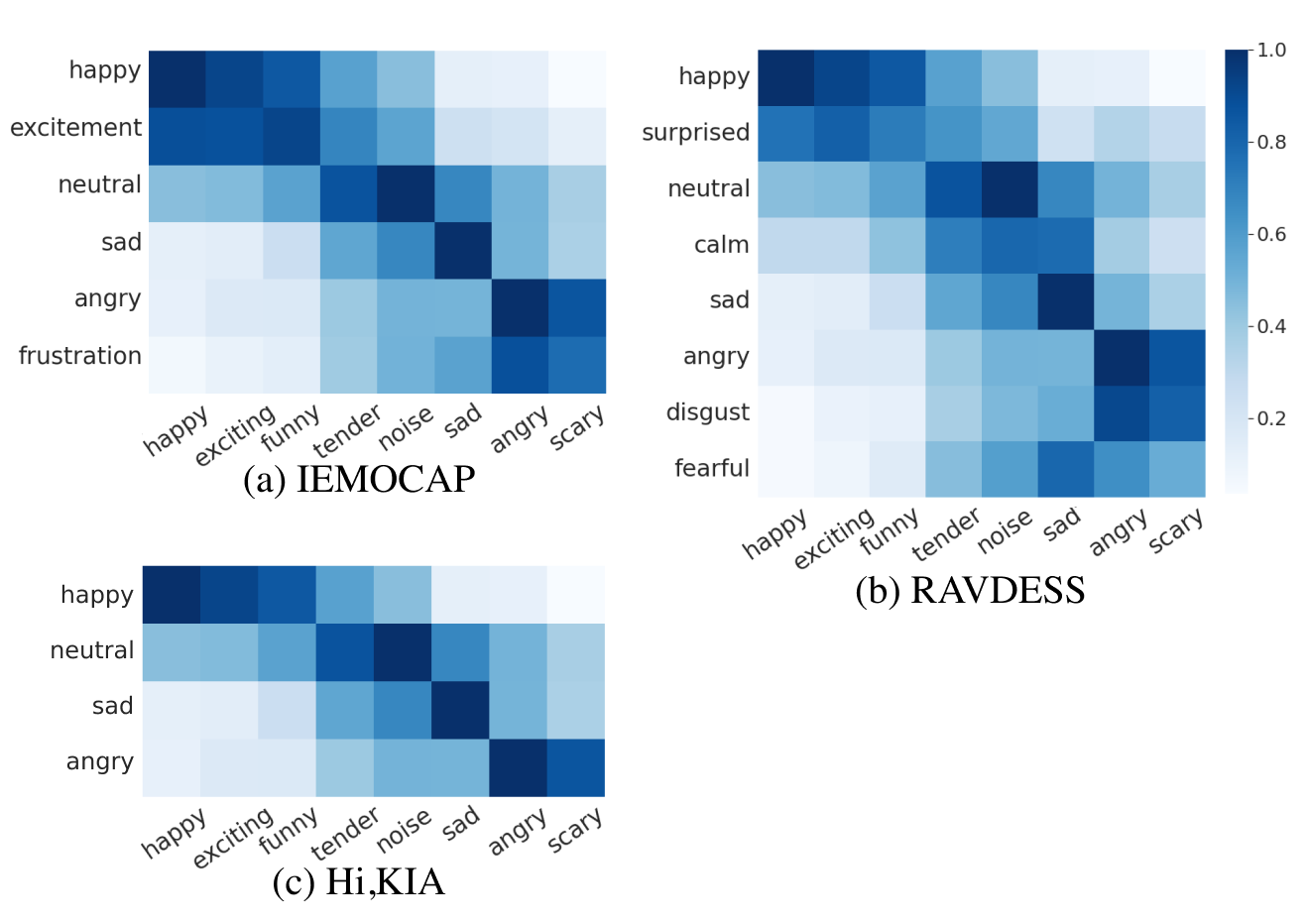}
\caption{Speech-to-Music Emotion Similarity using Valence Arousal Mapping. The x-axis is the Audioset music mood emotion label, and the y-axis is the emotion label of each speech dataset. The color is the value of valence-arousal similarity between the two labels.}
\label{fig:map}
\vspace{-2mm}
\end{figure}

Meanwhile, a different taxonomy problem occurs in the triplet sampling of $\mathcal{L}_{\text{Cross}}$. To solve this problem, we connect the emotion labels ($y_{s}, y_{m}$) of the two domains using valence-arousal mapping following the previous study \cite{won2021emotion}. After calculating the Euclidean distance between speech and music taxonomies (See Figure~\ref{fig:map}), we sample a music item that is the most similar to speech in the emotional space.

However, the triplet loss does not consider the neighborhood structure or data distribution within modalities. While the relationship between speech and music is considered, the relationship between each item and the emotion tag is not. The previous work~\cite{won2021emotion} addresses this problem with structure-preserving constraints~\cite{wang2016learning, kim2021learning}. With a ground truth emotion tag, each loss function is designed to optimize emotion-to-speech and emotion-to-music triplet loss as follows:
\begin{equation}
\begin{split}
\mathcal{L}_{\text{SP-Speech}} = \mathcal{L}_{\text{Triplet}}(z_{e},z_{s}^{+},z_{s}^{-}) \\
\mathcal{L}_{\text{SP-Music}} = \mathcal{L}_{\text{Triplet}}(z_{e},z_{m}^{+},z_{m}^{-})
\end{split}
\end{equation}
where $z_{e}$ denotes emotion tag embedding. We use the GloVe pre-trained word embedding~\cite{pennington2014glove} with a projection MLP layer for emotion tag embedding $z_{e}$. The model learns a shared embedding space between emotion, speech, and music. With these constraints, our final loss is given by $\mathcal{L}_{\text{Triplet+SP}}= w_{1} \mathcal{L}_{\text{Cross}} + w_{2} \mathcal{L}_{\text{SP-Speech}} + w_{3} \mathcal{L}_{\text{SP-Music}}$ where $w$ allows us to control the trade-off between the cross-domain mapping and the within-domain emotion mapping. Following a previous work \cite{won2021emotion}, we used 0.4, 0.3, and 0.3 for all $w_{1}, w_{2}, w_{3}$, respectively.

\subsection{Emotion Similarity Regularization}
To preserve the continuous emotion distribution in joint embedding space, we adopt the work of Gong~\textit{et al}, which proposed the use of rank similarity regularization (RankSim) \cite{gong2022ranksim} for deep imbalanced regression task. We propose the emotion similarity regularization (EmoSim), a modified version of RankSim, for the cross-domain retrieval task. An overview of our approach is shown in Figure \ref{fig:2}. In practice, our goal is to encourage the alignment between the similarity of neighbors in emotion space ($S^{y}$) and the similarity of neighbors in feature space ($S^{z}$). The EmoSim regularization term is formulated as follows:

\begin{equation} \label{eq:ranksim}
    \mathcal{L}_{\text{EmoSim}}= \frac{1}{N} \sum^{N}_{i=1}\ell(S^{y}_{i}, S^{z}_{i})
\end{equation}
where $S^{y}_{i}$ and $S^{z}_{i}$ denote the similarity scores of $i$ th speech and music items respectively, $\ell$ is a mean squared error, and $N$ is the mini-batch size. During training, we only use the unique similarity scores and enhance the relative representation of infrequent labels. Instead of calculating the rank, we directly compute the difference between the two similarity scores. With this regularization, our final loss is given by $\mathcal{L}_{\text{Triplet+Emosim}}= \mathcal{L}_{\text{Cross}} + \lambda \mathcal{L}_{\text{EmoSim}}$


where $\lambda$ allows us to control the trade-off between most similar emotion mapping (discrete) and overall emotion distribution mapping (continuous). The value of $\lambda$ was set to 0.5 when emotion regularization is applied.


\section{Experimental Design}
\subsection{Speech Datasets}
We validate the effectiveness of the proposed method on different types of speech datasets including acted speech (IEMOCAP), lexically-matched speech (RAVDESS), and wake-up word speech (HI,KIA). IEMOCAP~\cite{busso2008iemocap} contains approximately 12 hours of speech from 10 speakers and is divided into five sessions. Our experiment is based on 6 emotional classes. We use the first four sessions for training and the fifth session for the testing. We use 10\% of the provided training data as a validation set. RAVDESS~\cite{livingstone2018ryerson} is a multi-modal database of emotional speech and song. It features 24 different actors (12 males and 12 females) enacting 2 statements with 8 different emotions. We use the first 20 actors for training, actors 20-22 for validation, and actors 22-24 for test. Hi,KIA~\cite{kim2022hi} is a short-length wake-up word dataset labeled with four emotional states including anger, happy, sad, and neutral. It consists of 488 Korean accent utterances from 8 speakers. We use one of the official 8-fold splits, which shows the highest human validation performance (F7).

\subsection{Music Datasets}
We use AudioSet~\cite{gemmeke2017audio} that includes human-labeled 10-second sound clips drawn from YouTube videos. We use a subset under “music mood” ontology which contains music excerpts labeled with one of 7 music mood classes: \textit{happy, funny, sad, tender, exciting, angry,} and \textit{scary}. As some of the items are not available anymore, we actually use a smaller number of items than the official dataset includes. Because AudioSet~music~mood lacks neutral labels, we treat randomly generated color noise as a positive pair of neutral speech. The smaller dataset is provided with a training set of 18,643 clips (13,295 music, 5,348 noise) and an evaluation set of 405 clips (345 music, 60 noise).

\begin{table}[!t]
\centering
\resizebox{0.95\columnwidth}{!}{%
\begin{tabular}{l|cccc}
\toprule
Modality & AS Music Mood & IEMOCAP & RAVDESS & Hi,KIA \\ \midrule
Text & - & 0.709 & 0.133 & 0.250 \\
Audio & 0.605 & 0.718 & 0.733 & 0.935 \\ \bottomrule
\end{tabular}
}
\caption{Classification accuracy of music and speech emotion recognition models}
\label{tab:baseline}
\vspace{-2mm}
\end{table}

\begin{table*}[!t]
\centering
\resizebox{\textwidth}{!}{%
\begin{tabular}{lcclccclccclccc}
\toprule
 &  &  &  & \multicolumn{3}{c}{IEMOCAP \cite{busso2008iemocap}} &  & \multicolumn{3}{c}{RAVDESS \cite{livingstone2018ryerson}} &  & \multicolumn{3}{c}{HIKIA \cite{kim2022hi}} \\ \cmidrule{5-7} \cmidrule{9-11} \cmidrule{13-15} 
Method & Modality & Used In &  & MRR & P@5 & NDCG@5 &  & MRR & P@5 & NDCG@5 &  & MRR & P@5 & NDCG@5 \\ \midrule
Triplet & Text & \cite{won2021emotion} &  & 0.75±0.03 & 0.68±0.03 & 0.87±0.02 &  & 0.22±0.07 & 0.11±0.05 & 0.48±0.03 &  & 0.22±0.11 & 0.17±0.08 & 0.51±0.05 \\
Triplet + SP & Text & \cite{won2021emotion} &  & 0.76±0.01 & 0.68±0.02 & 0.88±0.01 &  & 0.29±0.03 & 0.16±0.06 & 0.54±0.06 &  & 0.23±0.14 & 0.16±0.1 & 0.52±0.09 \\
Triplet + EmoSim & Text & Ours &  & 0.76±0.01 & 0.69±0.02 & 0.88±0.01 &  & 0.28±0.1 & 0.18±0.05 & 0.62±0.06 &  & 0.22±0.11 & 0.1±0.03 & 0.46±0.02 \\ \midrule
Triplet & Audio & Ours &  & 0.74±0.02 & 0.67±0.04 & 0.86±0.03 &  & 0.72±0.03 & 0.65±0.03 & 0.85±0.03 &  & 0.84±0.09 & 0.73±0.06 & 0.91±0.03 \\
Triplet + SP & Audio & Ours &  & 0.73±0.03 & 0.65±0.02 & 0.86±0.02 &  & 0.75±0.04 & 0.65±0.05 & 0.87±0.02 &  & 0.83±0.10 & 0.73±0.10 & 0.88±0.06 \\
Triplet + EmoSim & Audio & Ours &  & 0.76±0.03 & 0.68±0.03 & 0.88±0.02 &  & 0.75±0.03 & \textbf{0.67±0.02} & \textbf{0.88±0.01} &  & \textbf{0.9±0.04} & \textbf{0.79±0.04} & \textbf{0.93±0.02} \\ \midrule
Triplet & Fusion & Ours &  & 0.80±0.03 & 0.73±0.05 & 0.89±0.02 &  & \textbf{0.76±0.02} & 0.65±0.02 & 0.86±0.01 &  & 0.87±0.06 & 0.74±0.10 & 0.92±0.04 \\
Triplet + SP & Fusion & Ours &  & 0.82±0.01 & 0.74±0.02 & 0.90±0.02 &  & 0.73±0.07 & 0.66±0.02 & 0.87±0.01 &  & 0.82±0.16 & 0.75±0.07 & 0.91±0.08 \\
Triplet + EmoSim & Fusion & Ours &  & \textbf{0.83±0.01} & \textbf{0.75±0.02} & \textbf{0.90±0.01} &  & 0.74±0.04 & 0.63±0.05 & 0.86±0.02 &  & 0.82±0.07 & 0.73±0.04 & 0.92±0.02 \\ \bottomrule
\end{tabular}
}
\caption{Speech-to-music retrieval results on three different speech datasets. \textbf{SP} stands for structure preserving, and \textbf{EmoSim} stands for emotion similarity regularization.}
\label{tab:retrieval}
\end{table*}

\subsection{Evaluation}
We use accuracy to evaluate the performance of the classification. For retrieval model evaluation, following the previous work \cite{won2021emotion}, we use two evaluation metrics: Precision at~5 (P@5) and Mean Reciprocal Rank (MRR). The relevant score is a binary value measured by whether \textit{query speech} and \textit{target music} share the same emotion semantic (with Figure~\ref{fig:map}). However, this metric has two problems in that the rank order is not considered, and it does not reflect the continuous emotion similarity. For example, \textit{tender} is more similar to \textit{sad} than \textit{angry}. To solve the above problem, we use normalizing Discounted Cumulative Gain at~5 (nDCG@5), which was designed for ranking tasks with more than one relevance level \cite{jarvelin2002cumulated}. We define the prediction score as the similarity between speech and music item; and ground truth as the valence-arousal (VA) similarity of the item's emotion label.

\subsection{Training Details}
The input to the music encoder is a 10-second audio excerpt at a 22,050~Hz sampling rate. It is converted to a log-scaled mel~spectrogram with 128 mel bins, 1024-point FFT with a hann window, and a 512 hop size. The inputs of the speech audio encoder and text encoder are 16,000~Hz waveforms up to 16 seconds and tokenized text with a max length of 64, respectively. In the case of the cross-domain retrieval model, all models are optimized using AdamW and use a 64-batch size. The models are trained with a learning rate of 1e-4 and a margin $\delta$ of 0.4.

\section{Results}
\subsection{Quantitative Results}
Table~\ref{tab:baseline} compares the baseline performance on classification tasks to evaluate feature extraction performance. As somewhat obviously, the audio modality works better on the lexically matched speech (`RAVDESS', `Hi,KIA'). The impressive part is that the audio modality shows high performance even in the IEMOCAP dataset with linguistic information. This shows that the audio modality is sufficient for emotion recognition without a speech-to-text model. 

Table~\ref{tab:retrieval} compares the different retrieval models with text, audio, fusion speech modalities. Each cross-domain retrieval model was trained 5 times with a different random seed and their average scores and standard deviations are reported. In the case of the fusion method, we utilize inter-modality dynamics through late fusion. \footnote{In the preliminary experiments, we found that the late fusion shows better performance than the early fusion} When comparing single modalities, the audio modality shows high performance in lexically matched speech datasets (`RAVDESS', `Hi,KIA'). In IEMOCAP with linguistic information, the text modality demonstrates comparable results to the audio modality. A possible reason might be the availability of high emotional clues in the linguistic structure. The fusion model shows a clear performance improvement over other single modalities. 

    
When comparing each method, our proposed emotion similarity (EmoSim) regularization method shows high performance and low standard deviation in all three datasets and various modalities. In particular, in the audio modality of the three datasets, \textit{triplet + EmoSim} method shows high NDCG@5 performance. This is because, in the failure cases, the model retrieves music items as close as possible to the correct answers.

\subsection{Qualitative Results}
\begin{figure}[!t]
\centering
\includegraphics[width=\columnwidth]{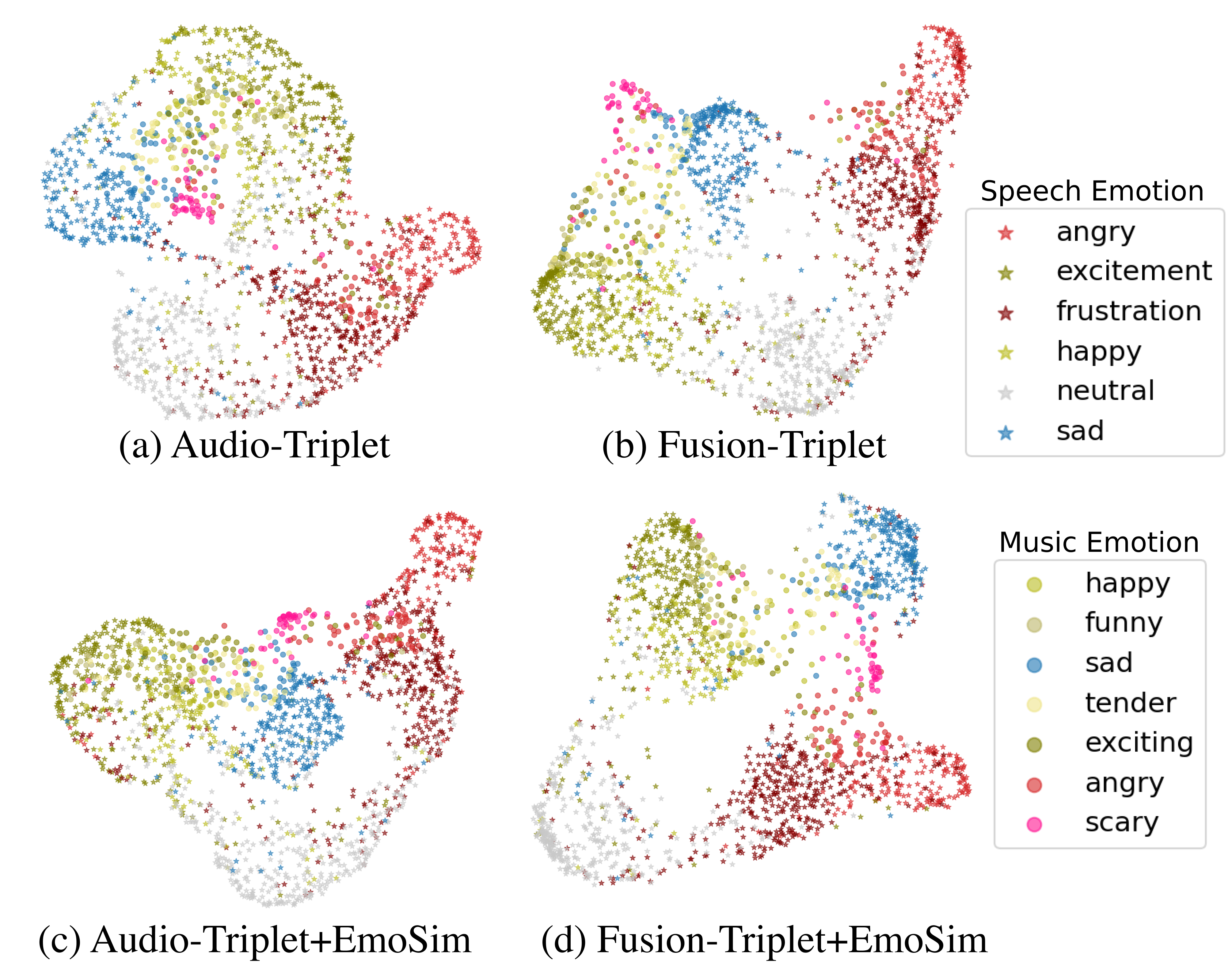}
\caption{UMAP visualization of speech-music joint embedding space. Star markers stand for speech embeddings and circle markers stand for music embedding.}
\label{fig:umap}
\vspace{-4mm}
\end{figure}

Metric learning embedding spaces are projected to a 2D space using uniform manifold approximation and projection (UMAP) \cite{mcinnes2018umap}. We fit UMAP with music and speech embeddings, then projected them into the 2D space. Figure~\ref{fig:umap} is the visualization of the multi-modal embedding space. Due to space constraints, we only visualize four models. All embedding vectors are computed from the test set. 

First, we observe that all models successfully discriminate emotion semantics. In addition, we can see a continuous distribution of emotions leading to \textit{happy$\xrightarrow{}$excited$\xrightarrow{}$tender$\xrightarrow{}$sad} in all models. However, in the case of the triplet model, \textit{scary} music embeddings are located between sad and happy speech embedding cluster (Figure~\ref{fig:umap} (a), (b)).  According to figure~\ref{fig:map}, in valence arousal space, scary emotion is similar to \textit{frustration} and \textit{angry} emotion. This problem is alleviated in \textit{Triplet + EmoSim} model. There are relatively few scary music samples closer to angry and frustration clusters (Figure~\ref{fig:umap} (c), (d)). We believe that joint embedding space learned inter-modality neighborhood structure from the continuous emotion similarity.

\section{Conclusions}
We presented a speech-to-music cross-domain retrieval framework that finds music that matches the emotion of speech. We explored various speech modality representations and proposed emotion similarity regularization term in cross-domain retrieval task. The experimental results demonstrated that, in the absence of linguistic information (lexically matched speech), the audio modality can retrieve music that matches the speaker's emotion. Especially, our proposed regularization term helps the joint embedding space understand the continuity of emotions. In the future, our approach can be extended to include other modalities (e.g., vision) and improved with better hierarchical fusion methods. 



\bibliographystyle{IEEEbib}
\bibliography{strings,refs}

\end{document}